\documentclass[twocolumn, secnumarabic, amssymb, amsmath, nobibnotes, aps, prl]{revtex4}
\usepackage{graphicx}
\usepackage{mathrsfs}
\usepackage{bm}

\begin{document}

\title{Diffusion on asymmetric fractal networks}
\author{Christophe P. Haynes and Anthony P. Roberts}
\affiliation{School of Mathematics and Physics, University of Queensland 4072, Australia}
\begin{abstract}
We derive a renormalization method to calculate the spectral dimension $\bar{d}$ of deterministic self-similar networks with arbitrary base units and branching constants. The generality of the method allows the affect of a multitude of microstructural details to be quantitatively investigated. In addition to providing new models for physical networks, the results allow precise tests of theories of diffusive transport. For example, the properties of a class of non-recurrent trees ($\bar{d}>2$) with asymmetric elements and branching violate the Alexander Orbach scaling law. 
\end{abstract}

\maketitle

The behavior of random walks on networks has been linked to a wide
range of interesting phenomena in many disciplines~\cite{Avra}.
In recent decades the study of the properties of fractal\cite{Havlin} and, more recently, complex networks~\cite{Condamin2,Gallos} has revealed a range of ``anomalous " behavior, characterized by power law scaling with non-integer exponents or dimensions.
In general, interactivity on networks can be codified in a matrix, which in many applications (e.g\ spring systems,
random walks and conductivity) takes a Laplacian form~\cite{Burioni5}. Macroscopic properties, such as phase-transitions~\cite{Avra,Burioni7},
and mass and electronic transport~\cite{Havlin,Benichou} are connected to the eigenspectrum of the matrix which is characterized by the
spectral dimension $\bar{d}$. The dimension $\bar{d}$ can be found using the site-independent~\cite{Hattori} asymptotic form of the probability $p_0(t) \sim t^{-\bar{d}/2}$ that a walker returns to its origin at time $t$. 

\begin{figure}
\begin{center}
\includegraphics[width = 0.46\textwidth]{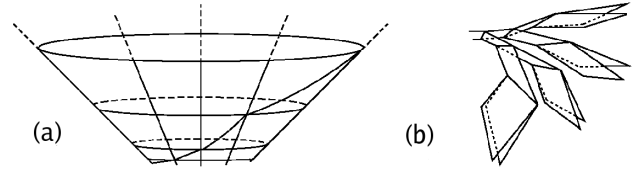}
\end{center}
\caption{Examples of networks. In (a) the base unit is a periodic comb with seven teeth and one diagonal strut. In (b) the network has branching constant $u=2$ and the base unit has the shape of an asymmetric staple.} 
\vspace{-5mm}

\label{fig:tree1}
\end{figure}

\begin{figure}
\begin{center}
\includegraphics[width = 0.45\textwidth]{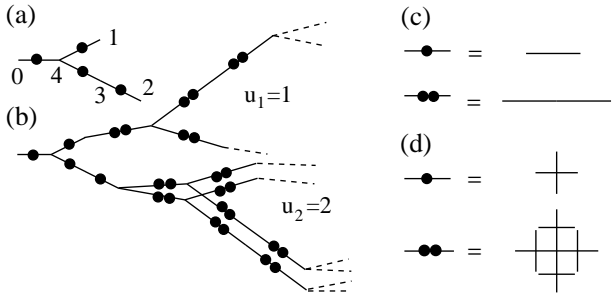}
\end{center}
\caption{ (a) The base unit, (b) the second iteration of the network is obtained by attaching multiple copies of the rescaled base unit to the end points of the base unit, (c) a regular pipe and (d) an iterated fractal.}
\label{fig:tree2}
\vspace{-5mm}

\end{figure}

Although many networks are random, simple models such as deterministic fractals, combs and trees have long been of theoretical interest because their properties can be analytically studied; the results used to test existing results, motivate new theories, mimic physical networks, or simply to provide insights into the nature of processes on networks. In this paper we use renormalization ideas~\cite{Rammal,Van2} to study $\bar{d}$ for classes of deterministic networks that include asymmetric elements and branching constants (See Figs.~\ref{fig:tree1} \&~\ref{fig:tree2}). The models may therefore be able to mimic asymmetric branching and chirality of polymers, as well as aspects of complex networks which exhibit high variability in their degree distribution~\cite{Albert}. 

In a seminal paper, Alexander and Orbach (AO)~\cite{Orbach} argued that $\bar{d}$ was related to the fractal dimension $d_f$ and random walk dimension
$d_w$ by the simple formula $\bar{d}=2d_f/d_w$. This result is one of the most important relations in the theory of fractals. All analytic results to date confirm the law, and
Telcs has proved that it holds for recurrent ($\bar{d}<2$) loopless networks which satisfy a technical smoothness condition on the electrostatic potential which arises in the definition of the network resistance. No proof has been given purely in terms of geometric and topological characteristics of networks. To demonstrate the renormalization method we calculate
$\bar{d}$, $d_f$ and $d_w$ for a class of networks which violates the AO law. In a related 
study~\cite{HRgfe09} we explain the discrepancy in terms of anisotropic diffusion on the network.

The method of iteratively constructing the infinite network is shown in Fig.~\ref{fig:tree2}. The tree is made by taking a base unit, doubling its size, and attaching $u_i$ copies of the re-scaled unit to each of the two end points of the base and so forth.
There are a number of generalizations which can be readily incorporated, allowing
the affect of numerous topological network characteristics to be probed.
First, the base unit can be an arbitrary sub-network with $L$ left, and $R$ right nodes (The tree example in 
Fig.~\ref{fig:tree2} has $L=1$ and $R=2$). If $L>R$, then $L-R$ of left nodes can be fused to form a network with $L=R$. If $L \leq R$ there are `$R$ choose $L$' distinct ways of connecting the first iteration to the base; a different branching constant $u_i$ can be associated with each possibility.
Second, each of the elements in the $n$th iteration of the tree can be replaced by the $n$th generation of a fractal with two specified end nodes (e.g. two corners of a Sierpinski triangle). 
Different sub-classes of these networks have been studied previously~\cite{Burioni2,Kron,Woess,Burioni3,Tony}.

In standard diffusion the probability that a random walker is at a point $x$ at time $t$ on a pipe is governed by the equation $p_t=Dp_{xx}$. To formulate the equations governing the probability on a network, we first solve the diffusion equation on a single bar for arbitrary Dirichlet conditions and a homogenous initial condition; $p \left( 0,t \right) = p_0\left(t\right)$, $p \left( b,t \right)  = p_1\left(t\right)$ and $p\left(x,0\right) = 0$.
The Laplace transform (LT) of the solution is~\cite{Haynes}
\begin{displaymath} 
c\left( x,s \right) = c_1(s)\,\frac{\sinh \left( x\sqrt {{\frac {s}{D}}}
 \right) }{\sinh \left(b\sqrt {{\frac {s}{D}}}\right)}+ c_0(s)\, \frac{\sinh \left(  \left(b-x \right) \sqrt {{
\frac {s}{D}}} \right) }{ \sinh \left(b\sqrt {{\frac {s}{D}}}\right)}.
\end{displaymath} 
The flux entering the bar at the node $x=0$ and $x=b$ is $q_0(t) = -Dp_x \left(0,t\right)$ and $q_1(t) = Dp_x\left(b,t\right)$ respectively. For networks it is useful to express these fluxes in terms of the concentrations $c_i$ at either end. This gives rise to the ``flux-concentration" equations
\begin{displaymath}
\frac{\textrm{tanh}(b\sqrt{s/D})}{\sqrt{sD}}\left[ \begin {array}{c} {j_0}\\\noalign{\medskip}{j_1}
\end {array} \right] =\left[ \begin {array}{cc} 1 & - \sigma
\\\noalign{\medskip}- \sigma & 1 \end {array} \right]  \left[ \begin {array}
{c} { c_0}\\\noalign{\medskip}{c_1}\end {array} \right],
\end{displaymath}
where $\sigma = \textrm{sech}(b\sqrt{s/D})$.
For simplicity we set $D$ to unity. In general, the bar can be replaced by the generator of a fractal (see Fig.~\ref{fig:tree2}(d)).  The above matrix is then a special case of the more general equation  
\begin{displaymath}
(1-f^2)g\left[ \begin {array}{c} {j_0}\\\noalign{\medskip}{j_1}
\end {array} \right] =\left[ \begin {array}{cc} 1 & - f
\\\noalign{\medskip}-f & 1 \end {array} \right]  \left[ \begin {array}
{c} { c_0}\\\noalign{\medskip}{c_1}\end {array} \right].
\end{displaymath}
Here $f$ is the LT of the first passage time density between the ends of the generator and $g$ is the LT of the probability that a walker released at the left end, is at its origin at time $t$. These functions are defined for a generator with reflective boundary conditions at either end~\cite[Appx.\ A]{Tony}. 
For a pipe, $f = \sigma$ and $g = (\textrm{tanh}(b\sqrt{s})\sqrt{s})^{-1}$ while for the generator of a determinstic tree $f = \sigma^2(4-3\sigma^2)^{-1}$ and $g = (4-3\sigma^2)(4(1-\sigma^2))^{-1}\textrm{tanh}(b\sqrt{s})/\sqrt{s}$~\cite[Appx.\ B]{Tony}. 

Using these relations, diffusion on an arbitrary network can be formulated as a system of algebraic equations~\cite{Haynes}. To link two or more pipes, the concentrations  at the common node are taken to be equal. Mass conservation at a node is enforced by requiring that the total flux entering all the pipes at that node sum to zero.  Using the node labels in Fig.~\ref{fig:tree2}(a), the equations for the example are 
\begin{displaymath}
(1-f^2)g {\bf j} =  \left[ \begin {array}{ccccc} 1&0&0&0&-f \\\noalign{\medskip}0&1&0& 0&-f
\\\noalign{\medskip}0&0&1&-f&0\\\noalign{\medskip}0& 0 &-f & 2 & -f 
 \\\noalign{\medskip} -f &-f &0 &-f &3\end {array} \right] {\bf c} ,
\end{displaymath} 
where ${\bf c}  = [c_{{0}}, c_{{1}}, c_{{2}}, c_{{3}}, c_{{4}}]^T$ {\em etc}.  
The number of equations for any sub-network with $L$ left end nodes, $R$ right end nodes and $K$ interior nodes can be reduced by setting $j_k = 0$ ($k = 1..K$) and solving the under determined system to give 
\begin{equation}\label{eq:cell8}
(1-f^2)g \left[ \begin{array}{c} J_{{L1}}\\\noalign{\medskip} J_{{R1}}\end {array} \right] = E\left[ \begin{array}{c} C_{{L1}}\\\noalign{\medskip}C_{{R1}}
\end {array} \right].
\end{equation}
Here $E$ is an $(L+R)\times (L+R)$ matrix and the subscripts in the above matrices denote their size (e.g. $J_{L1}$ is an L by 1 matrix). 

It is useful to rewrite equation \eqref{eq:cell8} as $\Gamma{\bf j} = {\bf c},$ where $\Gamma = (1-f^2)g E^{-1}$ is a matrix of Green's functions. The equation can be written in terms of sub-matrices as
\begin{equation}\label{eq:Tony1}
\left[ \begin {array}{cc} \Gamma_{{LL}} & \Gamma_{{LR}}\\\noalign{\medskip} \Gamma_{{
RL}}& \Gamma_{{RR}}\end {array} \right]
 \left[ \begin {array}{c} J_{{L1}}\\\noalign{\medskip} J_{{R1}}\end {array} \right]
 = \left[ \begin {array}{c} C_{{L1}}\\\noalign{\medskip}C_{{R1}}
\end {array} \right].
\end{equation}
Tree networks of the type shown in Fig.~\ref{fig:tree1} and~\ref{fig:tree2} can be iteratively constructed from these reduced networks.  Carrying out the reduction process for a network of arbitrary iteration gives a system of $L+R_{e}$ equations, where $R_{e}$ is the number of end nodes. If no-flux conditions are applied at these nodes ($j_{i} = 0$, $i = 1,\ldots,R_e$), the equations reduce to an $L\times L$ system. For the infinite network these equations are written as
\begin{equation}\label{eq:cell9}
\widehat{J}_{L1} = \frac{\sqrt{s}\widehat{E}(\textrm{sech}(b\sqrt{s}))}{\textrm{tanh}(b\sqrt{s})} \widehat{C}_{L1} = \widehat{\Gamma}^{-1}  \widehat{C}_{L1},
\end{equation} 
where the hat denotes a quantity associated with the infinite network.
To completely specify the problem, $L$ boundary conditions have to be applied at the left end nodes.
At the origin, which is defined to be the first of the $L$ nodes, an instantaneous unit source $q_0(t) = \delta(t)$ is applied
(so $j_0(s) = \mathscr{L}(q_0(t)) = 1$). No-flux conditions (i.e.\ $j_i = 0$) are applied to the remaining $L-1$ nodes. $p_0(t)$ is then given by the inverse LT of the first entry of the matrix $\widehat{\Gamma}$. If the origin is shifted, its concentration is given by the corresponding diagonal entry of $\widehat{\Gamma}$.  

Because of the self similar nature of the tree, the matrix $\widehat{\Gamma}$ can be studied through renormalization. The first step is to consider the equations for the infinite network which has been doubled in size. This is achieved by replacing $b$ in \eqref{eq:cell9} by $2b$ to get
\begin{equation}\label{eq:cell10}
 \widetilde{J}_{L1} = \frac{\sqrt{s}\widehat{E}(\textrm{sech}(2b\sqrt{s}))}{\textrm{tanh}(2b\sqrt{s})} \widetilde{C}_{L1} = \widetilde{\Gamma}^{-1} \widetilde{C} _{L1}. 
\end{equation}
The tilde is used to denote quantities associated with the rescaled network. Comparison between the two Green's functions in Eqns.~\eqref{eq:cell9} and~\eqref{eq:cell10} reveals that $\widetilde{\Gamma}(s) = 2\widehat{\Gamma}(4s)$.

The original network is reconstructed by patching the rescaled network to the initial finite network with $L$ left nodes and $R$ right nodes. If $L = R$ this is simply a matter of taking $J_{R1} +u\widetilde{J}_{L1}  = 0$, where $u$ is the branching constant (e.g., $u=2$ in Fig.~\ref{fig:tree1} (b)). For the example in Fig.~\ref{fig:tree2} ($R = 2, L = 1$), the mass conservation condition is 
\begin{equation}\label{Frr2}
J_{R1} + u_1\left[ \begin {array}{c} 0 \\\noalign{\medskip}\widetilde{J}_{{L1}}
\end {array} \right] + u_2\left[ \begin {array}{c} \widetilde{J}_{{L1}} \\\noalign{\medskip}0
\end {array} \right] = 0.
\end{equation}
In general, the conservation of mass condition can be written as 
\begin{equation}\label{Frr}
J_{R1} =  -\frac{1}{2}F_{RR}({\bf u},\widehat{\Gamma}^{-1}(4s))C_{R1},
\end{equation}
where ${\bf u} = (u_1,u_2,\ldots)$ are the branching constants. The form of $F_{RR}$ is dependent on the structural details of the connections. In general $F_{RR}$ depends on the elements of $\widehat{\Gamma}^{-1}$ individually. 

Now Eq.~\eqref{Frr} is used in the second equation of \eqref{eq:Tony1} to express $C_{R1}$ in terms of $J_{L1}$, which can then be used in the first equation of \eqref{eq:Tony1} to get $\Gamma^\dagger J_{L1} = C_{L1}$,
where $\Gamma^\dagger$ is calculated by algebraic methods from the terms above. The renormalization is completed by noting that $\Gamma^\dagger$ must, by construction, be identical to $\widehat{\Gamma}$ given in Eq.~\eqref{eq:cell9}. This leads to the closed equation
\begin{equation}\label{eq:Tony4}
\widehat{\Gamma}(s) = \Gamma_{{LL}} - \Gamma_{{LR}}\left(2F_{RR}^{-1}(u,\widehat{\Gamma}^{-1}(4s)) + \Gamma_{RR} \right)^{-1}\Gamma_{{RL}} .
\end{equation}
The result can be extended to the case where each pipe is replaced by a two-ended fractal. Subsequent generations of the network are defined by increasing the iteration of the fractal rather than doubling the length of the pipes. 
For an infinite network, Eq.~\eqref{eq:cell9} becomes 
\begin{equation}\label{eq:Tony5}
 \widehat{J}_{L1} = (g(1-f^2))^{-1}\widehat{E}(f) \widehat{C}_{L1} = \widehat{\Gamma}^{-1} \widehat{C}_{L1}.
\end{equation}
Increasing the generation of each fractal within the infinite network gives a structure whose equations (analogous to Eq.~\eqref{eq:cell10}) are
\begin{equation}\label{eq:Tony6}
\widetilde{J}_{L1} = (g^*(1-f^{*2}))^{-1}\widehat{E}(f^*) \widetilde{C}_{L1} = \widetilde{\Gamma}^{-1} \widetilde{C}_{L1}.
\end{equation}
Here $f^*$ and $g^*$ are the first passage time and Green's function associated with the two ends of the first iteration of the generator with reflective boundary conditions applied at both exterior nodes.  

The functions $f(s)$ and $f^*(s)$ are related by $f(\rho(s)) = f^*(s)$ where $\rho(s)$ is the renormalization of first passage time function~\cite{Van, Van2}.  For example, the deterministic tree has $\rho(s) = \textrm{arcsech}(\sigma^2/(4-3\sigma^2))^2$~\cite{Tony}. Applying this relation to the two infinite Green's function forms given in \eqref{eq:Tony5} and \eqref{eq:Tony6} gives $ \widetilde{\Gamma}(s) = g^*(s)\widehat{\Gamma}(\rho(s))/g(\rho(s)).$
Proceeding as above gives
\begin{equation}\label{renormbar}
\widehat{\Gamma} = \Gamma_{{LL}} - \Gamma_{{LR}}\left( \frac{g^*F_{RR}^{-1}(u,\widehat{\Gamma}(\rho(s))^{-1}))}{g(\rho(s))} + \Gamma_{RR} \right)^{-1}\Gamma_{{RL}}.
\end{equation}
Regarding the pipe as a fractal with $\rho(s) = 4s$ recovers the prior case [Eq.~\eqref{eq:Tony4}] as $g^*(s)/g(\rho(s)) = 2$ and $g^*(s) = (\textrm{tanh}(2b\sqrt{s})\sqrt{s})^{-1}$. Note that the renormalization of the tree with fractal elements is undertaken with respect to time, rather than space.

The spectral dimension $\bar{d}$ is found by considering 
the asymptotic behavior of the first element of the matrix $\widehat{\Gamma}$, denoted by $\hat{\gamma}$,
which has the form
\begin{equation}\label{eq:cell14}
\hat{\gamma}(s) = \left\{ \begin{array}{ccc}
k_{1}s^{\bar{d}/2-1}  + \ldots   & \bar{d} <2 \\
k_0s^{\bar{d}/2-1}\ln(s) + \ldots  &  \bar{d} = 2,4,6,\ldots \\
k_{0} + k_{1}s^{\bar{d}/2-1} + \ldots &   \bar{d} > 2, \bar{d} \neq 4,6,\ldots \\
\end{array} \right.
\end{equation}
The Green's function of the zeroth and first iteration have the asymptotic forms 
\begin{equation}\label{eq:cell15}
g = 1/(Ms) + O(s^0) \qquad g^* = 1/(M^*s) + O(s^0),
\end{equation}
where $M$ and $M^*$ are the mass of the zeroth and first iterations of the generator. The mass rescaling factor of the fractal
between successive iterations is given by $A=M^*/M$. For $s\to0$, $\rho(s)=a s +O (s^2)$, where
$a$ is the time rescaling factor. The ratio of the mean first passage times between the left and right end nodes of the $n-1$th and $n$th iterations of the generator is equal to $a$.

As an example, consider the infinite structure in Fig.~\ref{fig:tree2}(b). The matrix $F_{RR}$ required for the evaluation of Eq.~\eqref{renormbar} is determined (from Eq.~\eqref{Frr2}) to be 
\begin{eqnarray}\label{eq:cell25}
F_{RR} =  \frac{1}{\hat{\gamma}(\rho(s))} \left[ \begin {array}{cc} u_1 & 0 \\\noalign {\medskip} 0 & u_2 \end 
{array} \right].
\end{eqnarray}
The spectral dimension can be calculated by substituting \eqref{eq:cell25} into \eqref{renormbar} and then using the asymptotic forms \eqref{eq:cell14} and \eqref{eq:cell15}. There are several distinct cases. 
If $A (u_1 + u_2) - a \leq 0 $ then $\bar{d}\leq 2$ with $\bar{d} = 2\log_{a}(A(u_{1} + u_{2}))$,
while if  $A (u_1 + u_2) - a > 0 $ then $\bar{d}>2$ with 
\begin{eqnarray*}
\bar{d} = 2\,\log_a  \left( {\frac {A \left(  \left( u_{{1}}+u_{{2}} \right) x+6\,u
_{{2}}u_{{1}} \right) ^{2}}{4\,{u_{{1}}}^{2}u_{{2}}+ \left( 16\,{u_{{2
}}}^{2}+{x}^{2}+12\,xu_{{2}} \right) u_{{1}}+{x}^{2}u_{{2}}}} \right)  
\end{eqnarray*}
where $x>0$ is found by solving the quadratic equation
\begin{eqnarray*}
{x}^{2} \left( \left( u_{{1}}+u_{{2}} \right)A - a 
 \right) \\
+2\,x \left( 3Au_{{1}}u_{{2}}-a(3 u_{{2}}+ 2\,u_{{1}}) \right) -20\,u_{{2}}u_{{1}}a. 
\end{eqnarray*} 
Interestingly, the latter formula for $\bar{d}>2$ reduces to $\bar{d} = 2\log_{a}(A(u_{1} + u_{2}))$
when $u_1=2u_2$.

\begin{figure}
\includegraphics[width = 0.45\textwidth]{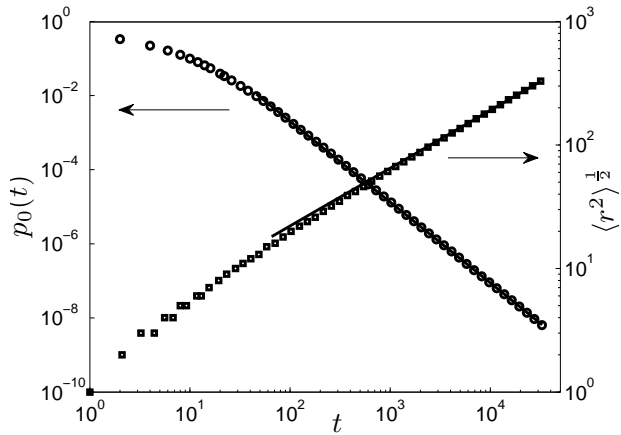}
\caption{Simulation of $\bar{d}$ and $d_w$. The left hand axis gives $p_0(t)$; we calculate $\bar{d} = 4.37$  (Best Fit: $4.36$). On the right hand axis we plot $\sqrt{r^{2}(t)}$ against $t^{1/2}$ confirming $d_w = 2$.}
\label{fig:fig2}
\vspace{-5mm}

\end{figure}

The AO law can be readily checked for a pipe ($A = 2$ and $a = 4$).
At large $t$ and $r$ the random walker's behavior will be dominated by diffusion on long pipes which implies $d_w  = 2$. A simple renormalization argument shows that the mass $M(r)$ of the network at a large distance $r$ from the origin is given by
$M(r) = 4 + 2 u_1 M((r-3)/2) + 2u_2 M((r-2)/2).$
Assuming $M(r) \sim  r^{d_f}$ and taking expansions gives $d_f = \ln(2(u_1 +u_2))/\ln(2)$. 
For $\bar{d}\leq2$, and for the exceptional case of $\bar{d}>2$ ($u_1=2u_2$), the pipe networks therefore obey the AO
law. Exluding the case $u_1=2u_2$ the results show that $\bar{d}\neq 2d_f/d_w$ if $\bar{d}>2$. 
Computations confirming the analytic values of $d_w$ and $\bar{d}$ for $u_1=1$ and $u_2=20$ are shown in Fig.~\ref{fig:fig2}. 
The properties of the tree network with a general fractal generator are qualitatively similar; it is apparent 
that the form $\bar{d} = 2\log_{a}(A(u_{1} + u_{2}))$ is just the AO law. 

Our main example demonstrates that networks with asymmetric characteristics can have properties which violate the AO law. Interestingly these unusual networks provide an example of a system in which local microscopic details affect macroscopic properties. This is not expected  in the scaling theory of dynamic phenomena; in general the network properties should be determined by power law exponents while microscopic details are relegated to pre-factors. This idea is demonstrated by considering the addition of a bar between nodes 1 and 3 of the tree shown in Fig.~\ref{fig:tree2}. As expected, this does not change $d_f$ or $d_w$ ($A$ and $a$ remain unchanged). In contrast, $\bar{d}$ is changed significantly and AO now holds if $u_1=4u_2$ when $\bar{d}>2$.

We have argued in Ref.~\cite{HRgfe09} that the AO law holds if each site a distance $r$ from the origin is explored approximately uniformly, and numerically demonstrated that this is not the case for $\bar{d}>2$. The fact that AO holds for the tree if $\bar{d}\leq 2$ demonstrates that asymmetry alone is not a sufficient condition for anisotropic transport for recurrent walks ($\bar{d} \leq 2$). The generality of the
renormalization method proposed should assist in the identification of geometric and topological characteristics associated with anisotropic (non AO-law) transport. In particular, the algebraic formulation allows a
number of results to be stated for arbitray sub-networks. For example, it is possible to prove~\cite{Tony2} that the AO law is obeyed for the class of networks considered here if $\bar{d} < 2$ or for $\bar{d}>2$ when $L=R=2$.

In Ref.~\cite{HRgfe09} it was shown that the resistivity exponent $\zeta$ of a network satisfies
the law $\zeta = \frac{d_w}{2}(2-\bar{d})$, which differs from the conventional 
relationship~\cite{Havlin} $\zeta=d_w - d_f$ if $\bar{d} \neq 2d_f/d_w$. A non-trivial test of the general formula is provided by replacing the pipe of the base unit by a deterministic tree generator ($A=4$, $a=8$). This gives $\bar{d}=3.58$, $d_w=3$ and hence $\zeta= -2.37$, which matches the exact value~\cite{HRgfe09}. 

We have derived a method for finding $\bar{d}$ for self-similar trees based on an arbitrary sub-network with fractal elements.  The ability to incorporate loops, branching and asymmetry allows the affects of these characteristics to be studied, and increases the classes of physical networks that can be quantitatively modeled. An advantage of using the continuum diffusion equation (compared with discrete-space walkers) is that non-integer bar lengths can be studied by taking $\rho(s)=4^vs$, with $v$ real and positive. This allows $\bar{d}$ to be continuously varied and careful study of the transition between recurrent and transient random
walks (at $\bar{d}=2$), as well as the transition from fractal to inhomogeneous trees ($v\to 0$) (e.g. the Bethe lattice).


\begin{thebibliography}{21}
\expandafter\ifx\csname natexlab\endcsname\relax\def\natexlab#1{#1}\fi
\expandafter\ifx\csname bibnamefont\endcsname\relax
  \def\bibnamefont#1{#1}\fi
\expandafter\ifx\csname bibfnamefont\endcsname\relax
  \def\bibfnamefont#1{#1}\fi
\expandafter\ifx\csname citenamefont\endcsname\relax
  \def\citenamefont#1{#1}\fi
\expandafter\ifx\csname url\endcsname\relax
  \def\url#1{\texttt{#1}}\fi
\expandafter\ifx\csname urlprefix\endcsname\relax\def\urlprefix{URL }\fi
\providecommand{\bibinfo}[2]{#2}
\providecommand{\eprint}[2][]{\url{#2}}

\bibitem[{\citenamefont{ben Avraham and Havlin}(2000)}]{Avra}
\bibinfo{author}{\bibfnamefont{D.}~\bibnamefont{ben Avraham}} \bibnamefont{and}
  \bibinfo{author}{\bibfnamefont{S.}~\bibnamefont{Havlin}},
  \emph{\bibinfo{title}{Diffusion and Reactions in Fractals and Disordered
  Systems}} (\bibinfo{publisher}{Cambridge Univ. Press},
  \bibinfo{address}{Cambridge, UK}, \bibinfo{year}{2000}).

\bibitem[{\citenamefont{Havlin and ben Avraham}(2002)}]{Havlin}
\bibinfo{author}{\bibfnamefont{S.}~\bibnamefont{Havlin}} \bibnamefont{and}
  \bibinfo{author}{\bibfnamefont{D.}~\bibnamefont{ben Avraham}},
  \bibinfo{journal}{Adv. Phys.} \textbf{\bibinfo{volume}{51}},
  \bibinfo{pages}{187} (\bibinfo{year}{2002}).

\bibitem[{\citenamefont{Condamin et~al.}(2007)\citenamefont{Condamin, Benichou,
  Tejedor, Voituriez, and Klafter}}]{Condamin2}
\bibinfo{author}{\bibfnamefont{S.}~\bibnamefont{Condamin}},
  \bibinfo{author}{\bibfnamefont{O.}~\bibnamefont{Benichou}},
  \bibinfo{author}{\bibfnamefont{V.}~\bibnamefont{Tejedor}},
  \bibinfo{author}{\bibfnamefont{R.}~\bibnamefont{Voituriez}},
  \bibnamefont{and} \bibinfo{author}{\bibfnamefont{J.}~\bibnamefont{Klafter}},
  \bibinfo{journal}{Nature} \textbf{\bibinfo{volume}{450}}, \bibinfo{pages}{77}
  (\bibinfo{year}{2007}).

\bibitem[{\citenamefont{Gallos et~al.}(2007)\citenamefont{Gallos, Song, Havlin,
  and Makse}}]{Gallos}
\bibinfo{author}{\bibfnamefont{L.~K.} \bibnamefont{Gallos}},
  \bibinfo{author}{\bibfnamefont{C.}~\bibnamefont{Song}},
  \bibinfo{author}{\bibfnamefont{S.}~\bibnamefont{Havlin}}, \bibnamefont{and}
  \bibinfo{author}{\bibfnamefont{H.~A.} \bibnamefont{Makse}},
  \bibinfo{journal}{Proc. Natl. Acad. Sci} \textbf{\bibinfo{volume}{104}},
  \bibinfo{pages}{7746} (\bibinfo{year}{2007}).

\bibitem[{\citenamefont{Burioni and Cassi}(2005)}]{Burioni5}
\bibinfo{author}{\bibfnamefont{R.}~\bibnamefont{Burioni}} \bibnamefont{and}
  \bibinfo{author}{\bibfnamefont{D.}~\bibnamefont{Cassi}}, \bibinfo{journal}{J.
  Phys. A: Math. Gen.} \textbf{\bibinfo{volume}{38}}, \bibinfo{pages}{R45}
  (\bibinfo{year}{2005}).

\bibitem[{\citenamefont{Burioni et~al.}(2006)\citenamefont{Burioni, Cassi,
  Corberi, and Vezzani}}]{Burioni7}
\bibinfo{author}{\bibfnamefont{R.}~\bibnamefont{Burioni}},
  \bibinfo{author}{\bibfnamefont{D.}~\bibnamefont{Cassi}},
  \bibinfo{author}{\bibfnamefont{F.}~\bibnamefont{Corberi}}, \bibnamefont{and}
  \bibinfo{author}{\bibfnamefont{A.}~\bibnamefont{Vezzani}},
  \bibinfo{journal}{Phys. Rev. Lett.} \textbf{\bibinfo{volume}{96}},
  \bibinfo{pages}{235701} (\bibinfo{year}{2006}).

\bibitem[{\citenamefont{Benichou and Voituriez}(2008)}]{Benichou}
\bibinfo{author}{\bibfnamefont{O.}~\bibnamefont{Benichou}} \bibnamefont{and}
  \bibinfo{author}{\bibfnamefont{R.}~\bibnamefont{Voituriez}},
  \bibinfo{journal}{Phys. Rev. Lett.} \textbf{\bibinfo{volume}{100}},
  \bibinfo{pages}{168105} (\bibinfo{year}{2008}).

\bibitem[{\citenamefont{Hattori et~al.}(1987)\citenamefont{Hattori, Hattori,
  and Watanabe}}]{Hattori}
\bibinfo{author}{\bibfnamefont{K.}~\bibnamefont{Hattori}},
  \bibinfo{author}{\bibfnamefont{T.}~\bibnamefont{Hattori}}, \bibnamefont{and}
  \bibinfo{author}{\bibfnamefont{H.}~\bibnamefont{Watanabe}},
  \bibinfo{journal}{Progr. Theoret. Phys. (Suppl.)}
  \textbf{\bibinfo{volume}{92}}, \bibinfo{pages}{108} (\bibinfo{year}{1987}).

\bibitem[{\citenamefont{Rammal}(1984)}]{Rammal}
\bibinfo{author}{\bibfnamefont{R.}~\bibnamefont{Rammal}}, \bibinfo{journal}{J.
  Stat. Phys.} \textbf{\bibinfo{volume}{36}}, \bibinfo{pages}{547}
  (\bibinfo{year}{1984}).

\bibitem[{\citenamefont{{Van den Broeck}}(1989{\natexlab{a}})}]{Van2}
\bibinfo{author}{\bibfnamefont{C.}~\bibnamefont{{Van den Broeck}}},
  \bibinfo{journal}{Phys. Rev. Lett.} \textbf{\bibinfo{volume}{62}},
  \bibinfo{pages}{1421} (\bibinfo{year}{1989}{\natexlab{a}}).

\bibitem[{\citenamefont{Albert and Barabási}(2002)}]{Albert}
\bibinfo{author}{\bibfnamefont{R.}~\bibnamefont{Albert}} \bibnamefont{and}
  \bibinfo{author}{\bibfnamefont{A.~L.} \bibnamefont{Barabási}},
  \bibinfo{journal}{Rev. Mod. Phys.} \textbf{\bibinfo{volume}{74}},
  \bibinfo{pages}{47} (\bibinfo{year}{2002}).

\bibitem[{\citenamefont{Alexander and Orbach}(1982)}]{Orbach}
\bibinfo{author}{\bibfnamefont{S.}~\bibnamefont{Alexander}} \bibnamefont{and}
  \bibinfo{author}{\bibfnamefont{R.}~\bibnamefont{Orbach}},
  \bibinfo{journal}{J. Phys. (Paris) Lett.} \textbf{\bibinfo{volume}{19}},
  \bibinfo{pages}{L625} (\bibinfo{year}{1982}).

\bibitem[{\citenamefont{Haynes and Roberts}(2009{\natexlab{a}})}]{HRgfe09}
\bibinfo{author}{\bibfnamefont{C.~P.} \bibnamefont{Haynes}} \bibnamefont{and}
  \bibinfo{author}{\bibfnamefont{A.~P.} \bibnamefont{Roberts}},
  \bibinfo{journal}{arXiv:0903.3279v1}  (\bibinfo{year}{2009}{\natexlab{a}}).

\bibitem[{\citenamefont{Burioni et~al.}(1998)\citenamefont{Burioni, Cassi,
  Pirati, and Regina}}]{Burioni2}
\bibinfo{author}{\bibfnamefont{R.}~\bibnamefont{Burioni}},
  \bibinfo{author}{\bibfnamefont{D.}~\bibnamefont{Cassi}},
  \bibinfo{author}{\bibfnamefont{A.}~\bibnamefont{Pirati}}, \bibnamefont{and}
  \bibinfo{author}{\bibfnamefont{S.}~\bibnamefont{Regina}},
  \bibinfo{journal}{J. Phys. A: Math. Gen.} \textbf{\bibinfo{volume}{31}},
  \bibinfo{pages}{5013} (\bibinfo{year}{1998}).

\bibitem[{\citenamefont{Kron and Teufl}(2004)}]{Kron}
\bibinfo{author}{\bibfnamefont{B.}~\bibnamefont{Kron}} \bibnamefont{and}
  \bibinfo{author}{\bibfnamefont{E.}~\bibnamefont{Teufl}},
  \bibinfo{journal}{Trans. Amer. Math. Soc.} \textbf{\bibinfo{volume}{356}},
  \bibinfo{pages}{393} (\bibinfo{year}{2004}).

\bibitem[{\citenamefont{Woess}(2000)}]{Woess}
\bibinfo{author}{\bibfnamefont{W.}~\bibnamefont{Woess}},
  \emph{\bibinfo{title}{Random walks on infinite graphs and groups}}
  (\bibinfo{publisher}{Cambridge University Press},
  \bibinfo{address}{Cambridge}, \bibinfo{year}{2000}).

\bibitem[{\citenamefont{Burioni and Cassi}(1995)}]{Burioni3}
\bibinfo{author}{\bibfnamefont{R.}~\bibnamefont{Burioni}} \bibnamefont{and}
  \bibinfo{author}{\bibfnamefont{D.}~\bibnamefont{Cassi}},
  \bibinfo{journal}{Phys. Rev. E} \textbf{\bibinfo{volume}{51}},
  \bibinfo{pages}{2865} (\bibinfo{year}{1995}).

\bibitem[{\citenamefont{Haynes and Roberts}(2009)}]{Tony}
\bibinfo{author}{\bibfnamefont{C.~P.} \bibnamefont{Haynes}} \bibnamefont{and}
  \bibinfo{author}{\bibfnamefont{A.~P.} \bibnamefont{Roberts}},
  \bibinfo{journal}{Phys. Rev. E} \textbf{\bibinfo{volume}{79}},
  \bibinfo{pages}{031111} (\bibinfo{year}{2009}).

\bibitem[{\citenamefont{Haynes and Roberts}(2008)}]{Haynes}
\bibinfo{author}{\bibfnamefont{C.~P.} \bibnamefont{Haynes}} \bibnamefont{and}
  \bibinfo{author}{\bibfnamefont{A.~P.} \bibnamefont{Roberts}},
  \bibinfo{journal}{Phys. Rev. E} \textbf{\bibinfo{volume}{78}},
  \bibinfo{pages}{041111} (\bibinfo{year}{2008}).

\bibitem[{\citenamefont{{Van den Broeck}}(1989{\natexlab{b}})}]{Van}
\bibinfo{author}{\bibfnamefont{C.}~\bibnamefont{{Van den Broeck}}},
  \bibinfo{journal}{Phys. Rev. A} \textbf{\bibinfo{volume}{40}},
  \bibinfo{pages}{7334} (\bibinfo{year}{1989}{\natexlab{b}}).

\bibitem[{\citenamefont{Haynes and Roberts}(0000{\natexlab{b}})}]{Tony2}
\bibinfo{author}{\bibfnamefont{C.~P.} \bibnamefont{Haynes}} \bibnamefont{and}
  \bibinfo{author}{\bibfnamefont{A.~P.} \bibnamefont{Roberts}},
  \bibinfo{journal}{Unpublished}  (\bibinfo{year}{0000}{\natexlab{b}}).

\end{thebibliography}

\end{document}